\documentclass[a4paper,12pt]{article}
\def\dalam{
\vbox{\hsize 0.5 em \hrule\hbox to 0.5 em{\vrule width 0.5 pt height 0.5 em
\hfill\vrule width 0.5 pt height 0.5 em} \hrule}  }

\title{The arions generation by magnetodipole 
waves\\ of pulsars and magnetars in a constant magnetic field}
\author{V.I.Denisov$^{1}$, G.A.Dantsev$^{1}$, V.I.Priclonsky$^{1}$,\\
 I.P.Denisova$^{2,3,a}$, O.N.Gavrish$^{2,3}$ \\
$^1$Department of Physics, Moscow State University, \\ 
119991, Moscow,  Russia, \\
$^{2}$ Bauman Moscow State Technical University, \\ 105005, Moscow,  Russia
\\
$^{3}$ The State University of Management, \\ 109542, Moscow, Russia
\\
$^{a}$ e-mail: pm.mati813@gmail.com (corresponding author)  \\ 
}
\begin{document}
\begin{titlepage}
\maketitle
\begin{abstract}
 The influence of the gravitational fields of pulsars and magnetars on the arion emission 
during the propagation of magnetodipole waves in a constant magnetic field has been evaluated.

The solution of the equation was obtained and the flux of arions emitted by magnetodipole 
waves during their propagation in a constant magnetic field was found. 
It is shown that the amplitude of the born arion wave at a distance from 
the source of magnetodipole radiation of a pulsar or magnetar $(r\to\infty)$ 
in the considered case tends to a constant value.
The intensity of the arion emission in the solid angle element and the amount of arion energy 
$\overline{I}$, 
emitted in all directions per unit time grow quadratically with increasing distance, 
traveled by the magnetodipole radiation of a pulsar or magnetar in a constant magnetic field.

Such growth of the energy of the born arion wave is due to the fact that in the considered 
problem constant magnetic field is defined in the whole space. In reality, the galactic and 
intergalactic magnetic fields can be represented in this form only in regions of space 
of finite dimensions, outside of which the force lines of their induction vector are curved. 
Therefore, it is possible to apply these results only in a region of space
for which $r\leq L_{coh}<\infty$, where $L_{coh}$ is the coherence length, the distance 
at which the force lines of the induction vector can be considered as straight lines.
An estimate for the value of the coupling constant of photons with arions is obtained.
\end{abstract}
\end{titlepage}
\section*{1. Introduction}
In the scientific literature of recent years, the processes of photoproduction of various 
axion-like particles beyond the Standart Model: arions [1,2] axions [3-6], and dilatons
 [7-10] are actively discussed.
These processes are currently regarded as the most realistic processes, 
with the help of which it is supposed [11-13] to carry out registration of axion-like 
particles in laboratory and astrophysical conditions.

The arion is a strictly massless pseudoscalar
Goldstone particle $a,$ which was introduced in 1982, in the papers [14-16] 
of Prof. A. A. Anselm and his co-authors.

The density of the Lagrangian function for the arion field, which is interacting with 
the electromagnetic field, is usually written in the canonical form:
$$L={\sqrt{-g}\over 2} g^{nm}{\partial a\over \partial x^n}                                   
{\partial a\over \partial x^m}-{\sqrt{-g}\over 16\pi}F_{nm}F^{nm}
-{g_{a\gamma}\sqrt{-g}\over4}  F_{nm}\tilde F^{nm} a,\eqno(1)
$$
where $a$ is the pseudoscalar field of the arion, 
$g_{a\gamma}$-- coupling constant of the arion with the electromagnetic field, $F_{nm}$ -- 
electromagnetic field tensor, $g$ is the determinant of the metric tensor,                                                      
$\tilde F^{nm}=E^{nmik}F_{ik}/2$ and $E^{nmik}=e^{nmik}/\sqrt{-g}$ -- the axial absolutely 
antisymmetric Levi-Civita tensor, and $e^{nmik}$ is the axial absolutely 
antisymmetric Levi-Civita symbol, and $e^{0123} =+1.$                 

When studying the processes of arion generation under astrophysical conditions
the influence of the gravitational field, in general, cannot be neglected.
Therefore, first of all, let us estimate the magnitude of this influence and the size of 
the region of space in which this influence can be significant.
Since the distribution of matter in neutron stars is close to spherically symmetric,
 then we will use the Schwarzschild solution as the metric tensor of the pseudo-Riemannian 
space. 
In the paper [18]  it is shown, that the Schwarzschild solution can be 
the external  solution for a non-spherically symmetric distribution of matter.

 The most convenient coordinates for writing down this solution are 
isotropic coordinates [17]. The nonzero components of the metric tensor in these
coordinates have the form:
$$g_{00}={(4r-r_g)^2\over (4r+r_g)^2},\ \ \ \ \ \
g_{xx}=g_{yy}=g_{zz}=-(1+{r_g\over 4r})^4,
 \eqno(2)$$
where $r_g$ is the Schwarzschild radius and the notation is used for convenience of reference: 
$r=\sqrt{x^2+y^2+z^2}.$
 
Let us estimate the value of the ratio $r_g/r$, included in expressions (2), on the surface of 
a neutron star. 
 The radius of a neutron star in recent times [19] is taken to be $R_s=10$ km, and the mass 
$M$ in the interval from 0.1 to 1.0 solar masses. 

Thus, in
our problem, the ratio $r_g/R_s\sim 0.01$. Since at $r>R_s$ the ratio $r_g/r$
takes smaller values, than the ratio $r_g/R_s$, the influence of the gravitational field 
is small and limited to a small neighborhood $r\leq 100 R_s$ of the neutron star. Therefore, 
as a first approximation for the small parameter $r_g/r$ in our problem as a metric tensor
we will use the metric tensor of the Minkowski space: 
$g_{00}=1,\ g_{11}=g_{22}=g_{33}=-1.$

The equations of the arion and electromagnetic fields derived from the Lagrangian density (1), 
in the Minkowski space have the form:
$$\dalam\ a=-{g_{a\gamma}} F_{mp}\tilde F^{mp}=-{g_{a\gamma}}({\bf B\ E}), \eqno(3)$$
$${\partial F^{nm}\over \partial x^m}=-4\pi \tilde F^{nm}{\partial a \over \partial x^m},$$
where ${\bf B}$ is the magnetic field induction vector, ${\bf E}$ is the electric field 
strength vector. 

According to the first equation (3), the source of arions are electromagnetic fields and waves, 
in which the first invariant of the electromagnetic field tensor is different from zero.
In the nature there exist such configurations of electromagnetic fields and waves at which 
this invariant is different from zero in large, by earthly standards, volumes of space.
These are, for example, 
magnetodipole radiation of pulsars and magnetars, propagating in the constant 
galactic or 
intergalactic magnetic field. And although the induction of galactic and intergalactic 
magnetic fields is relatively small at $B\sim 10^{-6}$ Gs,
the volumes occupied by these fields are significant. 
Therefore, it is of undoubted interest to study this process. Let us consider it in detail.
\section*{2. Calculation of arion emission arising from the propagation of magnetodipole waves 
of a pulsar or a magnetar in a constant magnetic field 
}
Suppose that at a point with radius vector ${\bf r} = {\bf r}_0=\{x_0,y_0,z_0\}$ 
is a pulsar or magnetar with a magnetic dipole moment $\bf m$, which rotates  
at a frequency of $\omega$ around an axis that makes an angle of $\psi$ with the vector $\bf m$.
Then this source emits an electromagnetic wave of frequency $\omega$, 
whose components [8] have the form:
$${\bf B}({\bf R},\tau)={3({\bf m}(\tau)\cdot {\bf R}){\bf R}
-R^2{\bf m}(\tau)\over R^5}-{{ \dot{\bf m}}(\tau)\over c R^2}+
\eqno(4)$$
$$+{3({ \dot{\bf m}}(\tau)\cdot {\bf R}){\bf R}\over c R^4}
+{(\ddot{\bf m}(\tau)\cdot {\bf R}){\bf R}-R^2\ddot{\bf m}(\tau)\over c^2 R^3},$$                              
$${\bf E}({\bf R},\tau)={({\bf R}\times\dot{\bf m}(\tau))\over c R^3}+
{({\bf R}\times\ddot{\bf m}(\tau))\over c^2 R^2},$$
where ${\bf R}={\bf r}-{\bf r}_0,$ the dot above the vector means the 
derivative on retarded time 
$\tau=t-R/c$, and the pulsar or the magnetar 
magnetic dipole moment in the following task has the components:
$${\bf m}(\tau)=|{\bf m}|\{\cos(\omega \tau)\sin\psi, \ 
\sin(\omega \tau)\sin\psi,\ \cos\psi\}.\eqno(5)$$
In the coordinate system, the origin of which is placed in the
point ${\bf r} = {\bf r}_0 $, the directional diagram of electromagnetic radiation (4) 
has the form
$$\overline{\left({dI\over d\Omega}\right)}_{EMW}={[{\bf r}\ddot{\bf m}]^2\over 4\pi c^3r^2} .$$
Integrating this expression over the angles $\theta$ and $\varphi$, we obtain the total 
intensity of radiation of a pulsar or a magnetar:
$$\overline{I}_{EMW}={2\ddot{\bf m}^2\over 3 c^3}={2c k^4{\bf m}^2\sin^2\psi\over 3}.\eqno(6)$$
As shown in [2,8], the electromagnetic wave (4) can emit arions and dilatons with 
frequencies     $\omega$ and $2\omega$.

Let there is a constant magnetic field in the considered region of space
$${\bf B}_0=\{B_{0x},B_{0y},B_{0z}\}.\eqno(7)$$
Substituting expressions (4) and (7) into equation (3) and considering that 
$${\bf E}({\bf r},\tau)={({\bf R}\times\dot{\bf m}(\tau))\over c R^3}+
{({\bf R}\times\ddot{\bf m}(\tau))\over c^2 R^2}=rot_{{\bf r}_0}{\dot{\bf m}(\tau)\over cR},$$
where $rot_{{\bf r}_0}$ is the rotor taking operator on coordinates of the vector 
${\bf r}_0=\{x_0,y_0,z_0\}$,
let's put it in the form
$$\dalam\ a=-{g_{a\gamma}}({\bf B}_0\ rot_{{\bf r}_0}{\dot{\bf m}(\tau)\over cR}). \eqno(8)$$
Now let's take into account that 
$$({\bf B}_0\ rot_{{\bf r}_0}{\dot{\bf m}(\tau)\over cR})=div_{{\bf r}_0}
\left[{\dot{\bf m}(\tau)\over cR}\ {\bf B}_0\right].$$

Introducing an auxiliary vector ${\bf F}$, we express the pseudoscalar field $a$ through it:
$$a=div_{{\bf r}_0}\left[{\bf F\ B}_0\right].$$

Then, to fulfill equation (8), it is necessary to require that this vector satisfies 
the equation:
$$\dalam\ {\bf F}=-{g_{a\gamma}}{\dot{\bf m}(\tau)\over cR}. \eqno(9)$$

Let's substitute expression (5) into the right part of this equation and for convenience of 
its solution we will write equation (9) in  complex form:
$$\dalam\ {\bf F}=-{{g_{a\gamma}}|{\bf m}|\omega\sin\psi\over c}\Big( i{\bf e}_x+{\bf e}_y\Big)
{\exp{i(\omega t-kR)}
\over R}. \eqno(10)$$ 
 After solving this equation, let's take only the real part.

We will look for the solution of equation (10) in the form:                
$$ {\bf F}=-{{g_{a\gamma}}|{\bf m}|\omega\sin\psi\over c}\Big( i{\bf e}_x+{\bf e}_y\Big)Q,$$
where the function $Q$
must satisfy the equation:
$$\dalam\ Q={\exp{i(\omega t-kR)}\over R}.$$
The retarded solution of this equation is of the form:
$$Q={i\over  2k}\exp{i(\omega t-kR)}. $$
As a result, for the real part of axion radiation arising at propagation of the 
magnetodipole 
radiation (4) of a pulsar or magnetar in a constant magnetic field (7), we have:
$$a={g_{a\gamma}|{\bf m}|k\sin\psi\over 2R}\Big\{B_{0z}\Big[(x-x_0)\cos[\omega t-kR]
+(y-y_0)\sin[\omega t-kR]\Big]-\eqno(11)$$
$$-(z-z_0)\Big[B_{0y}\sin[\omega t-kR]+B_{0x}\cos[\omega t-kR]\Big]\Big\}.$$
Let us pass to the Cartesian frame of reference, the origin of which is placed in the center 
of the pulsar or magnetar. 
Then $x_0=y_0=z_0=0$ and expression (11) will take the form:
$$a={g_{a\gamma}|{\bf m}|k\sin\psi\over 2r}\Big\{B_{0z}\Big[x\cos[\omega t-kr]
+y\sin[\omega t-kr]\Big]-\eqno(12)$$
$$-z\Big[B_{0y}\sin[\omega t-kr]+B_{0x}\cos[\omega t-kr] \Big]\Big\}.$$

Thus, the arion emission arising from the propagation of the magnetodipole radiation of a  
pulsar or magnetar radiation in a constant magnetic field occurs at the same frequency, 
as the frequency of the electric field that generates it.
\section*{3. Angular distribution of arion radiation}
Angular distribution of the arion emission,
 arising from the propagation of the electromagnetic wave (4) of a pulsar or magnetar 
through a permanent magnetic field (7), 
following the paper [8], we write it in the form:
$${dI\over d\Omega}=r\big({\bf r \ W}\big),\eqno(13)$$
where $\bf W$
is the energy flux density vector associated with
the components of the energy-momentum tensor $T^{nk}$
 by the relation $W^\beta =cT^{0\beta}$.
Using the expression for the energy-momentum tensor of the free arion field 
$$T^{nk}=g^{np}g^{km} \big\{{\partial a\over \partial x^p}
{\partial a\over \partial x^m}
-{1\over 2}g^{nk}{\partial a\over \partial x^p}
{\partial a\over \partial x^m} g^{pm}\big\},$$
from expression (13) we obtain:
$${dI\over d\Omega}=-cr^2{\partial a\over \partial r}{\partial a\over \partial x^0}.$$
Substituting into this relation the expression (12) 
for the arion field and leaving asymptotically the main 
part,
after time averaging, we reduce it to the form:
$$\overline{{dI\over d\Omega}}={g^2_{a\gamma}c|{\bf m}|^2k^4r^2\sin^2\psi\over 8}\Big\{
(B^2_{0x}+B^2_{0y})\cos^2\theta+B^2_{0z}\sin^2\theta-\eqno(14)$$
$$-2B_{0z}\sin\theta\cos\theta[B_{0x}\cos\varphi+B_{0y}\sin\varphi]\Big\}.$$
Let us now find the amount of arion energy $\overline{I}$ emitted in all directions per
 unit time:
$${\overline{I}}_{AR}=\int\limits_0^\pi \sin\theta d\theta\int\limits_0^{2\pi}d\varphi
 {\overline{ dI}\over d\Omega}={\pi g^2_{a\gamma}c|{\bf m}|^2k^4r^2\sin^2\psi\over 12}\Big\{
B^2_{0x}+B^2_{0y}+2B^2_{0z}\Big\}.\eqno(15)$$

The different dependence of this result on the components of the constant magnetic field (7) 
is due to the fact that the directivity diagram of the magnetodipole radiation of a pulsar 
or a magnetar is not spherically symmetric.
\section*{4. Conclusion}
It follows from expression (12) that the amplitude of the born arion wave at a distance from 
the magnetodipole source of the pulsar or magnetar $(r\to\infty)$ 
in the considered case tends to a constant value.
The intensity of the arion emission in the solid angle element and the amount of arion energy 
$\overline{I}$, 
emitted in all directions per unit time grow quadratically with increasing distance, traveled 
by the magnetodipole radiation of a pulsar or magnetar in a constant magnetic field (7).

Such growth of the energy of the born arion wave is due to the fact that in our problem 
the constant magnetic field (7)
is set in the whole space. In reality, the galactic and intergalactic magnetic fields 
can be represented in the form (7) only in regions of finite dimensions, outside of which 
the force lines of their 
induction vectors 
are curved. Therefore, one can apply the results (12), (14) and (15) only in the region 
of space,
for which $r\leq L_{coh}<\infty$, where $L_{coh}$ is the coherence length -- the distance 
at which the lines of force 
of the induction vector can be considered as straight lines.

Let us define the conversion factor of electromagnetic radiation energy into arion energy 
$\beta$
as the ratio of the intensity of radiation of arions ${\overline{I}}_{ar}$ to the intensity 
of the electromagnetic field
of a pulsar or magnetar ${\overline{I}}_{EMW}$:
$$\beta={{\overline{I}}_{AR}\over {\overline{I}}_{EMW}}
={\pi L^2_{cog}g^2_{a\gamma}\over 8}\Big\{
B^2_{0x}+B^2_{0y}+2B^2_{0z}\Big\}.$$
 
This coefficient reflects the properties of the converter, i.e., the electromagnetic field. 
It should be noted, 
that according to the second equation of the system (3), along with the conversion 
of the energy of electromagnetic waves 
into the energy of arions, the opposite process takes place. If $\beta << 1,$ 
then the reverse process  can be ignored; if the coefficient of $\beta$
is close to unity, it is necessary to consider both processes together.

At present the value of the photon-arion coupling constant $g_{a\gamma}$ is unknown.
Let's roughly estimate the value of this constant using the example of the generation of arions 
by the electric field of pulsar radiation as it propagates through the magnetic field of our 
Galaxy. 

According to modern data [20,21], the radius of the Galaxy is 16 kiloparsecs.
It is accepted to distinguish the large-scale component of the magnetic field of the Galaxy 
magnetic field of the Galaxy (the scale of homogeneity of the order of hundreds and thousands 
of parsecs) and 
 fluctuation component with a wide range of scales (from fractions of parsecs to hundreds 
of parsecs).
The induction of the large-scale magnetic field of the Galaxy is estimated to be 2-3 $\mu$ Gs.

Based on these data, it is reasonable to put $L_{coh}\sim 10^3$ ps $=3\cdot 10^{21}$ cm, 
$B_0\sim 10^{-6}$ Gs.

Then from the condition $\beta << 1$ we obtain:
$ g^2_{a\gamma}<8\cdot10^{-31}$ ${cm\over erg}.$
Equation (15) was derived by us into the Gaussian system of units of measurement.
Let us rewrite this formula in the natural system of
units. If we take into account that 1 erg= 624 GeV,
and 1 cm = $0.5 \cdot 10^{14}$ GeV$^{-1}$,
then we get: 
$g_{a\gamma}<0.9\cdot 10^{-10}$ GeV$^{-1}$.
This estimate coincides with the estimates of the coupling constant $g_{a\gamma}$  
obtained in [2,22-25].
\section*{Acknowledgements}
This study was conducted within the scientific program of the National Center for Physics and 
Mathematics, section $ \#5$ $<<$Particle Physics and Cosmology$>>.$ Stage 2023-2025.

\end{document}